\pdfoutput=1
\documentclass[12pt]{article}
\usepackage{cite}
\usepackage{graphicx}
\usepackage{latexsym}
\usepackage{mathrsfs}
\usepackage[overload]{textcase}

\font\grande=cmr9.5 scaled \magstep4
\font\medio=cmr9.5 scaled \magstep2
\outer\def\beginsection#1\par{\medbreak\bigskip
      \message{#1}\leftline{\bf#1}\nobreak\medskip
\vskip-\parskip
      \noindent}

\begin{document}
\bibliographystyle{unsrt}

\titlepage

\vspace{1cm}
\begin{center}
{\grande The scaling of primordial gauge fields}\\
\vspace{1.5 cm}
Massimo Giovannini \footnote{e-mail address: massimo.giovannini@cern.ch}\\
\vspace{1cm}
{{\sl Department of Physics, CERN, 1211 Geneva 23, Switzerland }}\\
\vspace{0.5cm}
{{\sl INFN, Section of Milan-Bicocca, 20126 Milan, Italy}}
\vspace*{1cm}
\end{center}
\vskip 0.3cm
\centerline{\medio  Abstract}
\vskip 0.5cm
The large-scale magnetic fields arising from the quantum mechanical fluctuations of the hypercharge are investigated when the evolution of the gauge coupling is combined with a sufficiently long inflationary stage. In this framework the travelling waves associated with the quantum mechanical initial conditions turn asymptotically into standing waves which are the gauge analog of the Sakharov oscillations.  Even if the rate of dilution of  the hypermagnetic and hyperelectric fields seems to be superficially smaller than expected from the covariant conservation of the energy-momentum tensor,  the standard evolution for wavelengths larger than the Hubble radius fully accounts for this anomalous scaling which is anyway unable to increase the amplitude of the magnetic power spectra after symmetry breaking. An effective amplification of the gauge power spectra may instead occur when the post-inflationary expansion rate is slower than radiation. We stress that the modulations of the gauge power spectra freeze as soon as the relevant wavelengths reenter the Hubble radius and not at the end of inflation. After the Mpc scale crosses the comoving Hubble radius the scaling of the magnetic power spectrum  follows from the dominance of the conductivity.  
From these two observations the late-time values of the magnetic power spectra are accurately computed in the case of a nearly scale-invariant slope and contrasted with the situation where the phases of Sakharov oscillations are not evaluated at horizon crossing but at the end of inflation, i.e. when all the wavelengths relevant for magnetogenesis are still larger than the comoving horizon. 

\noindent
\vspace{5mm}
\vfill
\newpage
The problem of the origin of large-scale magnetism has dates back to the pioneering contributions of Hoyle, Zeldovich, Harrison  \cite{hoyle,zeldovich,harrison} and many others. As correctly emphasized by Hoyle \cite{hoyle} one of the most mysterious aspects of the problem 
is related to the largeness of the correlation scale associated with the 
magnetic fields at the onset of the gravitational collapse of the protogalaxy.
This comoving scale should be as large as ${\mathcal O}(\mathrm{Mpc})$ 
and this means that it crossed the comoving horizon at a relatively late time, 
i.e. much later than electron-positron annihilation and just before 
matter-radiation equality. In the past it has been argued that if we combine the correlation scale of the protogalactic field with the inflationary dynamics  
we might hope that  the gauge fields are amplified from the inflationary vacuum. The problem with this assumption is that, besides the invariance under local gauge transformations, the Weyl \cite{lich} and the duality \cite{duality1,duality2} symmetries determine the evolution of the gauge fields in general relativity. When the governing equations of a given quantum field are invariant under Weyl rescaling, the  normal modes are not excited by the evolution of the geometry \cite{parker1,birrell,parker2,bcf0} and the corresponding particles are not produced. Consequently in conformally flat background geometries $\overline{G}_{\mu\nu} = a^2(\tau) \eta_{\mu\nu}$ (where $a(\tau)$ is the scale factor, $\eta_{\mu\nu}$ is the Minkowski metric and $\tau$ is the conformal time coordinate) the energy density of the hypercharge fields $\rho_{Y} = u_{\mu} \, u_{\nu} T_{Y}^{\mu\, \nu}$ scales as $a^{-4}$ and it is progressively diluted as the Universe expands.

 If the Weyl invariance is broken because of the coupling either to the inflaton or to some other spectator field \cite{bcf0,bcf1,bcf2,bcf3,bcf5,bcf6,bcf7,bcf8} the gauge modes can be amplified so that the quantum mechanical initial data (corresponding to travelling waves) turn asymptotically into standing waves whose phases only depend on the evolution of the gauge coupling. These modulations of the power spectra are the gauge analog of the so-called Sakharov oscillations \cite{SO1,SO2,SO3} and their correct evaluation determines the amplitude of the power spectra. The purpose of this analysis is twofold. In the first place we want to demonstrate that the late-time values of the magnetic power spectrum (and of the corresponding Sakharov oscillations) are not fixed at radiation dominance (and independently of the wavenumber) but they are instead determined by the moment when the relevant wavelengths (comparable with the Mpc) get of the order of the Hubble radius before equality. The second point involves the post-inflationary evolution since any deviation from the radiation dominance affects the late-time amplitude of the magnetic power spectrum at the gravitational collapse of the protogalaxy. These two themes are non-trivially combined and their 
 discussion is also quite confusing in the current literature.  Even if these two problems can be analyzed in even more general terms, from now on the attention will be focused on the simplest possible action, namely
\begin{equation}
S_{gauge} = - \frac{1}{4} \int d^{4} x \,\, \frac{\sqrt{- G}}{g^2} \, \, Y_{\mu\nu} \, Y^{\mu\nu}, \qquad\qquad 
g = \sqrt{\frac{4 \pi}{\lambda}},
\label{ac1}
\end{equation}
where $g$ denotes throughout the gauge coupling while $G$ is the determinant 
of the four-dimensional metric with signature mostly minus. Equation (\ref{ac1}) may be complemented by a pseudo-scalar coupling with relevant implications \cite{bcf1,bcf2} that are however not central to the present discussion. 
We stress however that the action (\ref{ac1}) involves hypercharge fields so that 
the electromagnetic degrees of freedom will have to be related to $Y_{\mu\nu}$ by the correct projections involving the Weinberg angle. For the record the energy-momentum tensor following from Eq. (\ref{ac1}) is
\begin{equation}
T_{Y}^{\mu\nu} = \frac{\lambda}{4 \pi} \biggl( - Y^{\mu\alpha} \, Y^{\nu}_{\,\,\,\,\alpha} + 
\frac{G^{\mu\nu}}{4} \,\, Y^{\alpha\beta} \,\, Y_{\alpha\beta} \biggr),
\label{ac2}
\end{equation}
where $\lambda$ is the susceptibility which is related to the gauge coupling as already mentioned in Eq. (\ref{ac1}). In spite of the amplification mechanism 
 Eq. (\ref{ac2})  suggests that energy density still scales $a^{-4}$ in terms of the comoving fields $\vec{E}$ and $\vec{B}$: 
\begin{equation}
\rho_{Y} = u_{\mu} \, u_{\nu} T_{Y}^{\mu\, \nu} = \frac{1}{8 \pi a^4}  \biggl( E^2 + B^2\biggr) = \frac{\lambda}{8 \pi}\biggl(b^2 + e^2\biggr).
\label{NUM1}
\end{equation}
In Eq. (\ref{NUM1}) the comoving fields $\vec{E} = a^2 \,\, \sqrt{\lambda} \vec{e}$ and $ \vec{B} = a^2 \sqrt{\lambda} \,\,\vec{b}$ have been introduced in terms of the  the physical fields $\vec{e}$ and $\vec{b}$. We note that in the conformally flat background considered here the physical fields $\vec{e}$ and $\vec{b}$ are directly related to the components of the gauge field strengths that are defined as $Y^{i\, 0} = e^{i}/a^2$ and $Y^{i\, j} = \, -\, \epsilon^{i \, j\, k} b_{k}/a^2$. 

If the background geometry is conformally flat (as it is the case during an inflationary stage of expansion) the evolution of the comoving fields only depends on the rate of variation of $g$ (or of $\sqrt{\lambda}$). In the absence of sources the fields $\vec{E}$ and $\vec{B}$ are both solenoidal and their dynamics follows from:
\begin{equation}
\vec{E}^{\,\prime} +  {\mathcal F}\, \vec{E} =  \vec{\nabla} \times \vec{B}, \qquad \qquad 
\vec{B}^{\,\prime}  -  {\mathcal F}\, \vec{B} = - \vec{\nabla} \times \vec{E},
\label{NUM2}
\end{equation}
where the prime denotes a derivation with respect to the conformal time coordinate $\tau$ and ${\mathcal F} = \sqrt{\lambda}^{\prime}/\sqrt{\lambda}= - g^{\prime}/g$ is the rate of variation of $\sqrt{\lambda}$ (or of $g$). The duality symmetry \cite{duality1,duality2} transforms the hyperelectric into the hypermagnetic fields and vice-versa\footnote{The duality symmetry implies that for $\sqrt{\lambda} \to 1/\sqrt{\lambda}$ (i.e. ${\mathcal F} \to - {\mathcal F}$) the comoving fields change as $\vec{E} \to \vec{B}$ and $\vec{B} \to - \vec{E}$ and the two expressions of Eq. (\ref{NUM2}) are transformed one into the other. We stress that the duality symmetry \cite{duality1} is explicit in terms of the comoving normal modes of the system \cite{duality2} and this is why their use is recommended for the class of problems discussed in this paper.}. Since ${\mathcal F} = \sqrt{\lambda}^{\prime}/\sqrt{\lambda} \equiv -g^{\prime}/g$  we have that ${\mathcal F} <0$ corresponds to the situation 
where the gauge coupling {\em increases} (while the susceptibility decreases at the same rate ); conversely when ${\mathcal F}>0$ the gauge coupling {\em decreases} (and the susceptibility increases at the same rate).

According to general arguments \cite{bcf9,bcf10} the inflationary stage must be long enough to iron any preexisting inhomogeneity but also sufficiently short to exclude a Planckian regime of initial curvatures. For the present purposes the total number of $e$-folds will then be taken between $60$ and $80$ so that the quantum mechanical initial conditions can be safely imposed for all the modes of the gauge field spectrum. In this framework the comoving fields of Eq. (\ref{NUM2}) are promoted to the status of field operators:
\begin{eqnarray}
\hat{E}_{i}(\vec{x}, \tau) &=& - \int\frac{d^{3} k}{(2\pi)^{3/2}} \sum_{\alpha} \hat{e}^{(\alpha)}_{i}(\hat{k}) \, 
\biggl[ g_{k}(\tau) \, \hat{a}_{\vec{k},\,\alpha} e^{- i \vec{k} \cdot\vec{x}}  + \mathrm{H}.\mathrm{c}.\biggr],
\label{NUM3}\\
\hat{B}_{k}(\vec{x}, \tau) &=& - \epsilon_{i\, j\,k} \int\frac{d^{3} k}{(2\pi)^{3/2}} \sum_{\alpha} k_{i} \, \,\hat{e}^{(\alpha)}_{j}(\hat{k}) \, 
\biggl[ f_{k}(\tau) \, \hat{a}_{\vec{k},\,\alpha} e^{- i \vec{k} \cdot\vec{x}}  - \mathrm{H}.\mathrm{c}.\biggr],
\label{NUM4}
\end{eqnarray}
where $[\hat{a}_{\vec{k},\alpha}, \hat{a}_{\vec{p}, \, \beta}^{\dagger} ] = \delta^{(3)}(\vec{k} - \vec{p})$ and $e^{(\alpha)}_{i}(\hat{k})$ (with $\alpha=\oplus,\,\,\otimes$) accounts for the two polarizations which are orthogonal to the direction of propagation $\hat{k}= \vec{k}/k$ (i.e. $\hat{e}^{\oplus} \times \hat{e}^{\otimes} = \hat{k}$). The quantum mechanical normalization demands that the Wronskian of the mode functions is time-independent and normalized as $f_{k}(\tau) g_{k}^{\ast}(\tau) - f_{k}^{\ast}(\tau) g_{k}(\tau) = i$; in this way the canonical commutation relations between field operators are preserved. All in all, provided the inflationary phase is sufficiently long, the gauge fields are initially 
in the vacuum and the expectation values of the field operators in Fourier space can be computed from  Eqs. (\ref{NUM3})--(\ref{NUM4}):
\begin{eqnarray}
&& \langle \hat{E}_{i}(\vec{k},\tau) \,  \hat{E}_{j}(\vec{p},\tau) \rangle = \frac{2\pi^2}{k^3} \, p_{i\, j}(\hat{k}) \,\, P_{E}(k, \tau) \,\, \delta^{(3)}(\vec{k} + \vec{p}), 
\label{NUM5}\\
&&\langle \hat{B}_{i}(\vec{k},\tau) \,  \hat{B}_{j}(\vec{p},\tau) \rangle = \frac{2\pi^2}{k^3} \, p_{i\, j}(\hat{k}) \,\,P_{B}(k, \tau) \,\, \delta^{(3)}(\vec{k} + \vec{p}), 
\label{NUM6}
\end{eqnarray}
where $p_{i\, j}(\hat{k}) = (\delta_{i\,j} - \hat{k}_{i} \hat{k}_{j})$. We stress that the hyperelectric and hypermagnetic power spectra (i.e. $P_{E}(k,\tau)$ and $P_{B}(k,\tau)$) appearing in  Eqs. (\ref{NUM3})--(\ref{NUM6}) have the dimensions of an energy density and their explicit expressions are:
\begin{equation}
P_{E}(k,\tau) = \frac{k^{3}}{2 \pi^2} \bigl| g_{k}(\tau) \bigr|^2, \qquad \qquad P_{B}(k,\tau) = \frac{k^{5}}{2 \pi^2} \bigl| f_{k}(\tau) \bigr|^2.
\label{NUM7}
\end{equation}
The evolution of the mode functions entering Eq. (\ref{NUM7}) follows by inserting Eqs. (\ref{NUM3})--(\ref{NUM4}) into Eq. (\ref{NUM2}) and by imposing the appropriate Wronskian normalization.
The gauge power spectra only depend on the evolution of ${\mathcal F}$, i.e. the rate of variation of the gauge coupling (or of the susceptibility). During inflation $a \, H = {\mathcal H}= - 1/[\tau \, (1- \epsilon)]$ where $\epsilon = - \dot{H}/H^2\ll 1 $ is the standard slow-roll parameter\footnote{ For quantitative estimates the validity of the consistency relations is also assumed and this means that $r_{T} = 16 \epsilon$; since according to current bounds \cite{TS1,TS2,TS3} $r_{T} < 0.06$ (or even $r_{T} < 0.036$), the 
upper limit on $\epsilon$ ranges between $3.7 \times 10^{-3}$ and  
$2.2 \times 10^{-3}$.}.

We could now consider a rather general evolution of the gauge coupling; however 
there are two (dual) inflationary backgrounds where the gauge power spectra acquire a particularly concise expression i.e. ${\mathcal F} = \pm 2 {\mathcal H}$ \cite{bcf11}. Furthermore, in these two cases the r\^ole played by the Sakharov oscillations is particularly transparent. If the gauge coupling {\em increases} during an inflationary stage {\em and then flattens out} we are led to consider the situation where, during inflation, ${\mathcal F}= - 2 {\mathcal H}$ while, after the end of inflation, $|{\mathcal F}/{\mathcal H}| \ll 1$. It is relevant that the transition across the inflationary boundary occurs in a way that $a$, $g$, ${\mathcal H}$ and ${\mathcal F}$ are all continuous; this can be achieved by keeping ${\mathcal F}/{\mathcal H}$ arbitrarily small (but always finite). If we impose the quantum mechanical initial 
data the mode functions behave asymptotically for $\tau \to - \infty$ as travelling waves
\begin{equation}
f_{k} = e^{- i \, k\tau}/\sqrt{2 k}, \qquad\qquad g_{k}(\tau) = - i \,  e^{- i\,k \tau}\,\sqrt{k/2},
\end{equation}
and using then the continuity of the mode functions across the inflationary boundary the gauge power spectra after the end of inflation follow from Eq. (\ref{NUM7}):
\begin{equation}
P_{E}(k, \tau) = \frac{9 H_{1}^4 \, a_{1}^4}{4 \pi^2} \cos^2{k\tau}, \qquad \qquad P_{B}(k, \tau) = \frac{9 H_{1}^4 \, a_{1}^4}{4 \pi^2} \sin^2{k\tau}.
\label{NUM8}
\end{equation}
Concerning Eq. (\ref{NUM8}) the following two comments are in order:
\begin{itemize}
\item{} $H_{1}$ is fixed by the value of tensor to scalar ratio and by the amplitude of the power spectrum of curvature inhomogeneities (i.e. ${\mathcal A}_{{\mathcal R}} = {\mathcal O}(2.41) \times 10^{-9}$)
according to $(H_{1}/M_{P}) = \sqrt{\pi \, {\mathcal A}_{{\mathcal R}} \, r_{T}}/4$;
\item{} even if the power spectra of Eq. (\ref{NUM8}) have been determined by positing that the gauge coupling $g \propto1/\sqrt{\lambda}$ {\em increases} during inflation and then flattens out later on, by using the duality symmetry of the problem \cite{duality1,duality2} 
(see Eq. (\ref{NUM2}) and discussion thereafter) we can deduce the power spectra in the dual situation where the gauge coupling first {\em decreases} as ${\mathcal F} = 2 {\mathcal H}$ during inflation and then flattens out; as expected from duality \cite{duality2} the spectra of Eq. (\ref{NUM8}) are therefore interchanged as:
\begin{equation}
\overline{P}_{E}(k, \tau) = \frac{9 H_{1}^4 \, a_{1}^4}{4 \pi^2} \sin^2{k\tau}, \qquad \qquad \overline{P}_{B}(k, \tau) = \frac{9 H_{1}^4 \, a_{1}^4}{4 \pi^2} \cos^2{k\tau};
\label{NUM9}
\end{equation}
\item{} in what follows we regard as more physical the case of increasing gauge coupling given in Eq. (\ref{NUM8}) however the same considerations can also be applied to Eq. (\ref{NUM9}).
\end{itemize}
The modulations appearing in 
Eqs. (\ref{NUM8})--(\ref{NUM9}) are the gauge analog of Sakharov oscillations \cite{SO1,SO2}; this analogy can be made even more concrete but and this discussion can be found in \cite{SO3}. It is 
interesting to stress, in this respect, that the duality symmetry interchanges the phases of the Sakharov 
oscillations since $P_{E}(k, \tau) \to \overline{P}_{B}(k, \tau)$ and $P_{B}(k, \tau) \to \overline{P}_{E}(k, \tau)$. The Sakharov phases are {\em generic} 
since they only depend on the rate of variation of the gauge coupling 
during inflation and by the requirement that, after the end of inflation, 
the gauge coupling flattens out (i.e. $|{\mathcal F}/{\mathcal H}| \ll 1$). If the gauge coupling {\em does not} freeze we could have the situation where $|{\mathcal F}/{\mathcal H}| {\mathcal O}(1)$ after the end of inflation; in this case the Sakharov phases are always present and their analytical expressions are given in terms of the appropriate Bessel functions \cite{SO3} whose trigonometric limit of Eqs. (\ref{NUM8})--(\ref{NUM9}) is realized exactly when $|{\mathcal F}/{\mathcal H}| \ll 1$.

We are now going to ask the following question: given the expression of the comoving power spectra of Eq. 
(\ref{NUM8}) what is the value of the {\em physical} power spectra at late time?
The answer follows from the relation between ($\vec{e}$, $\vec{b}$) and  ($\vec{E}$, $\vec{B}$) introduced in Eq. (\ref{NUM1}). In particular, from the general form of $P_{E}(k,\tau)$ and $P_{B}(k,\tau)$ (see Eq. (\ref{NUM7})) the relation between the physical and the comoving spectra is therefore given by:
\begin{equation}
{\mathcal P}_{X}(k,\tau) = \frac{P_{X}(k,\tau)}{\lambda(\tau) \, a^4(\tau)} = \frac{g^2(\tau)}{4 \, \pi\, a^4(\tau)}  \, \, P_{X}(k,\tau),\qquad\qquad X= E,\, B,
\label{NUM10}
\end{equation}
where the explicit connection  between the gauge coupling and the susceptibility (see Eq. (\ref{ac1}) and discussion thereafter) has been used. Consequently, if Eq. (\ref{NUM8}) is inserted into Eq. (\ref{NUM10}) the corresponding physical spectra become: 
\begin{equation}
{\mathcal P}_{E}(k,\tau) =\frac{9\, H_{1}^4 \, g_{1}^2}{16 \pi^3} \,\, \biggl(\frac{a_{1}}{a}\biggr)^4\,\,\cos^2{k\tau},
\qquad \qquad {\mathcal P}_{B}(k,\tau) =\frac{9\, H_{1}^4 \, g_{1}^2}{16 \pi^3} \,\, \biggl(\frac{a_{1}}{a}\biggr)^4\,\,\sin^2{k\tau}.
\label{NUM11}
\end{equation}
In Eq. (\ref{NUM11}) $g_{1}$ is the asymptotic value of the gauge coupling after the end of inflation
since for $\tau \gg -\tau_{1}$ the evolution of $g(\tau)$ has been parametrized as $g(\tau) \simeq g_{1} (\tau/\tau_{1})^{\delta}$ (with $\delta \ll 1$).  Moreover for $\tau < - \tau_{1}$ we have that $a H = -1/[(1-\epsilon)\tau]$ while for  $\tau \gg -\tau_{1}$ we may consider that the background is dominated by radiation\footnote{This choice is not mandatory and is relaxed later on. However, already at this stage it is useful to stress that the typical phases of the standing oscillations do not depend on the post-inflationary evolution of the scale factor but are determined by the asymptotic behaviour of the gauge coupling.}.  Equation (\ref{NUM11}) only applies  for typical time-scales shorter than the crossing time $\tau_{k} = {\mathcal O}(1/k)$ (i.e. in the range $-\tau_{1} < \tau \leq \tau_{k}$) so that when $\tau < \tau_{k}$ the standing oscillations can be expanded in their small argument limit:
\begin{eqnarray}
{\mathcal P}_{E}(k,\tau) &=& \frac{9\, H_{1}^4 \, g_{1}^2}{16 \pi^3} \,\, \biggl(\frac{a_{1}}{a}\biggr)^4\,\, \biggl[ 1 + {\mathcal O}\biggl(\frac{k^2}{{\mathcal H}^2}\biggr)\biggr],\qquad \qquad \tau < \tau_{k},
\label{NUM12}\\
{\mathcal P}_{B}(k,\tau) &=& \frac{9\, H_{1}^4 \, g_{1}^2}{16 \pi^3} \,\, \biggl(\frac{a_{1}}{a}\biggr)^4\,\, \biggl(\frac{k}{{\mathcal H}}\biggr)^2 \biggl[ 1 + {\mathcal O}\biggl(\frac{k^2}{{\mathcal H}^2}\biggr)\biggr], \qquad \qquad \tau < \tau_{k}.
\label{NUM13}
\end{eqnarray}
We can now rephrase the leading order result of Eq. (\ref{NUM13}) by recalling that ${\mathcal H} = a \, H$. The approximate scaling of the magnetic power spectra
is then given by the following pair of (fully equivalent) expressions
\begin{eqnarray}
{\mathcal P}_{B}(k,\tau) &=& \frac{9\, H_{1}^4 \, g_{1}^2}{16 \pi^3} \,\, \biggl(\frac{k}{a_{1} \, H_{1}} \biggr)^2\,\, 
\biggl(\frac{a_{1}}{a}\biggr)^6 \biggl(\frac{H_{1}}{H}\biggr)^2, 
\qquad \qquad \tau < \tau_{k},
\nonumber\\
&=& \frac{9\, H_{1}^4 \, g_{1}^2}{16 \pi^3} \,\, \biggl(\frac{a_{1}}{a}\biggr)^4 |k \, \tau|^2 , \qquad\qquad k \tau< 1.
\label{NUM14}
\end{eqnarray}
Even if the two interchangeable results of Eq. (\ref{NUM14}) coincide, by looking at the first line we could naively infer that between $\tau_{1}$ and the present time (be it $\tau_{0}$) the physical power spectrum does not scale as 
as $a^{-4}$ but rather as $a^{-6} H^{-2}$. This would mean, for instance, that during radiation (where $H \propto a^{-2}$) the power spectrum, as a whole, scales as ${\mathcal P}_{B}(k,\tau) \propto a^{-2}$. This conclusion is however misleading in the light of the second form of Eq. (\ref{NUM14}) suggesting instead that {\em there is no gain in the amplitude} and to clarify this point even further we now specifically focus on the first expression of Eq. (\ref{NUM14}) and note that the potential increase of the physical power spectrum is controlled by the last pair of terms:
\begin{equation}
 \biggl(\frac{k}{a_{1} \, H_{1}} \biggr)^2\,\, 
\biggl(\frac{a_{1}}{a}\biggr)^6 \biggl(\frac{H_{1}}{H}\biggr)^2 = |k \tau_{1}|^2 \biggl(\frac{a_{1}}{a}\biggr)^6 \biggl(\frac{H_{1}}{H}\biggr)^2.
\label{NUM14a}
\end{equation}
Even if, for some reason, $(a_{1}/a)^6 (H_{1}/H)^2$ gets much larger than $a^{-4}$, $|k \tau_{1}|$ is always so minute to jeopardize any potential growth of the first term. In particular we have that for the fiducial values of the parameters (determined in the framework of the concordance 
paradigm) $|k \tau_{1}|^2 = {\mathcal O}(10^{-46})$ (see also, in this respect, Eq. (\ref{NUM23}) and discussion thereafter). In other words, {\em in spite of any gain possibly coming from $(a_{1}/a)^6 (H_{1}/H)^2$  the potential advantages of the modified redshift are compensated by the smallness of $|k\tau_{1}|^2$}. Moreover, since by definition $k\tau <1$,  the physical power spectrum of Eq. (\ref{NUM14}) for $\tau < \tau_{k}$ {\em is always suppressed in comparison to its value at $\tau_{k}$.}:
\begin{equation}
{\mathcal P}_{B}(k,\tau) \ll \frac{9\, H_{1}^4 \, g_{1}^2}{16 \pi^3} \,\, \biggl(\frac{a_{1}}{a_{k}}\biggr)^4, \qquad\qquad \tau \leq \tau_{k}.
\label{NUM15}
\end{equation}
At most the whole expression in Eq. (\ref{NUM14a}) is ${\mathcal O}(1)$ 
and this happens when $ \tau= \tau_{k}$. All in all the situation can be summarized as follows:
\begin{itemize}
\item{} the hypermagnetic power spectrum, for any given time-scale $ \tau < \tau_{k}$, 
is always smaller than its value for $\tau \to \tau_{k}$, i.e. at horizon crossing;  
\item{} taking into account this 
effect the authors of Ref. \cite{bcf13} suggest a correction of previous estimates by $37$ orders of magnitude that would solve most of the problems of magnetogenesis since the rate of dilution of the magnetic fields would be much smaller than previously considered; this argument has been further amplified later on \cite{bcf14,bcf15}; 
\item{} in the light of Eqs. (\ref{NUM15})--(\ref{NUM16}) it is however difficult to see how the anomalous scaling can affect 
the physical value of the magnetic field for $\tau> \tau_{k}$ where the potential dilution is not effective \cite{bcf5,bcf6,bcf7,bcf8} (see also  \cite{bcf16a,bcf16,bcf17a,bcf17});
\item{} finally, neither Eq. (\ref{NUM13}) nor Eq. (\ref{NUM14}) apply for $\tau > \tau_{k}$: indeed as long as $ k \tau>1$, the electric fields are strongly damped in comparison with the magnetic ones which are also suppressed for typical wavenumbers shorter than the so-called magnetic diffusivity scale \cite{MOF,PAR,bcf18}.
\end{itemize}

The discussion of the previous paragraph boils down to the following pair of complementary questions: 
where should we evaluate the phases of the (magnetic) Sakharov oscillation appearing in Eq. (\ref{NUM11})? At  $\tau_{k}$ (i.e. when the Mpc scale crosses the Hubble radius) or at $\tau_{1}$ (i.e. at 
the end of inflation)? In what follows we shall consider separately the two cases and then conclude that the purported dilution effect only applies beyond the Hubble radius and has no appreciable effect on the final value of the magnetic field. To achieve this program it is useful to remark that 
the freezing time of the Sakharov phases cannot be selected at wish: the result of Eq. (\ref{NUM11}) applies for $\tau \leq \tau_{k}$ since, after $\tau_{k}$, the conductivity breaks the duality symmetry \cite{duality2} and the relevant evolution equations {\em are not given anymore}
by Eq. (\ref{NUM2}). This is why the results of Eqs. (\ref{NUM11})--(\ref{NUM13}) 
evaluated for  $\tau = {\mathcal O}(\tau_{k})$ 
\begin{equation}
{\mathcal P}^{(em)}_{B}(k,\tau_{k}) =\frac{9\, H_{1}^4 \, g_{1}^2}{16 \pi^3} \,\,\cos^2{\theta_{W}}\,\, \biggl(\frac{a_{1}}{a_{k}}\biggr)^4\,\,\sin^2{k\tau_{k}},
\label{NUM16}
\end{equation}
are subsequently scaled down to the epoch of the gravitational collapse of the protogalaxy.
For the sake of accuracy, in Eq. (\ref{NUM16}) we added the cosine squared of the Weinberg angle (coming from the projection of the hypercharge field on the electromagnetic field after symmetry breaking) so that ${\mathcal P}^{(em)}_{B}(k,\tau_{k})$ 
really estimates the {\em magnetic power spectrum} at $\tau_{k}$. The specific value of $\tau_{k}$ depends on the bunch of wavelengths relevant for magnetogenesis (i.e. $k = {\mathcal O}(\mathrm{Mpc}^{-1})$); this means that we can estimate the crossing time as $\tau_{k} = {\mathcal O}(10^{-2}) \,\, \tau_{eq}$ or, more precisely\footnote{As usual $\Omega_{M0}$ is the present matter fraction in critical units while $\Omega_{R0}$ is the radiation fraction; $h_{0}$, as usual, is the Hubble rate in units $\mathrm{km}\,\, \mathrm{Hz}/\mathrm{Mpc}$.}
\begin{equation}
\frac{\tau_{k}}{\tau_{eq}} = 1.01 \times 10^{-2} \biggl(\frac{k}{\mathrm{Mpc}^{-1}}\biggr)^{-1} 
\biggl(\frac{h_{0}^2 \Omega_{M0}}{0.1386}\biggr) \biggl(\frac{h_{0}^2 \Omega_{R0}}{4.15\times 10^{-5}}\biggr)^{-1/2}.
\label{NUM17}
\end{equation}
where $\tau_{eq}$ is the equality time. When  the comoving 
wavelength ${\mathcal O}(\mathrm{Mpc})$ is inside the Hubble radius for $\tau >\tau_{k}$ the duality symmetry relating Eqs. (\ref{NUM10})--(\ref{NUM11}) is broken by the presence of the conductivity that suppresses both the electric and the magnetic fields \cite{duality1,duality2} even if the electric fields are comparatively far more suppressed. If $\vec{{\mathcal B}}$ denotes the (ordinary) magnetic field
in the conducting plasma prior to equality, the Ohmic electric field is 
$\vec{{\mathcal E}} = \vec{\nabla} \times \vec{{\mathcal B}}/\sigma_{c}$ where $\sigma_{c}$ is the (comoving) conductivity \cite{MOF,PAR,bcf18}. This remark implies that the electric power spectrum is suppressed by a factor $(k/\sigma_{c})^2$ in comparison with its magnetic counterpart for $\tau \geq \tau_{k}$ and, in the same regime, the magnetic field is practically not suppressed for typical wavenumbers smaller than the magnetic diffusivity scale $k_{\sigma} = \sqrt{\sigma_{c} \, {\mathcal H}}$. All in all, at the present time and for large wavelengths (i.e. $k< k_{\sigma}$), the magnetic power spectrum is given by: 
\begin{equation}
{\mathcal P}^{(em)}_{B}(k, \tau) = {\mathcal P}^{(em)}_{B}(k, \tau_{k}) \, \, \biggl(\frac{a_{k}}{a}\biggr)^4, \qquad \qquad \tau > \tau_{k}.
\label{NUM18}
\end{equation}
If we now insert Eq. (\ref{NUM15}) into Eq. (\ref{NUM17}) and evaluate 
the spectrum at the present time we obtain from Eq. (\ref{NUM18}):
\begin{equation}
{\mathcal P}^{(em)}_{B}(k, \tau_{0}) = \frac{9\, H_{1}^4 \, g_{1}^2}{16 \pi^3} \,\,\cos^2{\theta_{W}}\,\, \biggl(\frac{a_{1}}{a_{0}}\biggr)^4 \sin^2{k \, \tau_{k}},
\label{NUM19}
\end{equation}
where, by definition, $ k \tau_{k} = {\mathcal O}(1)$. The estimate of the term $(a_{1}/a_{0})$ appearing in Eq. (\ref{NUM19}) depends on the post-inflationary expansion history and, in the simplest case, we have $(a_{1}/a_{0}) = (2 \, \Omega_{R0})^{1/4} \sqrt{H_{0}/H_{1}}$ where $\Omega_{R0}$ is the present fraction of relativistic species and $H_{0}$ is the Hubble rate. We may then recall that the overall normalization of the temperature and polarization anisotropies fixes the value of $H_{1}$ in terms of the tensor to scalar ratio $r_{T}$ \cite{TS1,TS2,TS3}. Equation (\ref{NUM19}) then takes the general form 
\begin{equation}
{\mathcal P}^{(em)}_{B}(k,\tau_{0}) =  \frac{9\,\, g_{1}^2\,\, \cos^2{\theta_{W}}}{128 \, \pi^2} \,\, \Omega_{R0}\,\, H_{0}^2 \, M_{P}^2\, {\mathcal A}_{{\mathcal R}}\, \,r_{T} \,\,\sin^2{k \tau_{k}}.
 \label{NUM20}
 \end{equation}
If we now recall  the fiducial values of the quantities appearing in Eq. (\ref{NUM20}) we can deduce the following estimate of the magnetic power spectrum:
 \begin{equation}
 \frac{{\mathcal P}^{(em)}_{B}(k,\tau_{0})}{\mathrm{nG}^2} \,= 3.39\times 10^{- 8}\,\,\biggl(\frac{r_{T}}{0.06}\biggr)\,\, \biggl(\frac{g_{1}}{0.1}\biggr)^2\,\,
 \biggl(\frac{{\mathcal A}_{{\mathcal R}}}{2.41 \times 10^{-9}}\biggr)\,\, \biggl(\frac{h_{0}^2 \, \Omega_{R\,0}}{4.15 \times 10^{-5}}\biggr),
 \label{NUM21}
 \end{equation}
 where $1\,\, \mathrm{nG} = 10^{-9} \mathrm{G}$; we also remind that , in natural units, the relation 
 between the Planck mass and the Gauss is given by $M_{P}^2 = 10^{57.33} \,\, \mathrm{G}$. Finally in Eq. (\ref{NUM21} we took into account that the phase of the standing oscillation must be evaluated for $\tau \to \tau_{k}$ so that  $\sin{k \, \tau_{k}} = {\mathcal O}(0.84)$ since $k\, \tau_{k} = {\mathcal O}(1)$. 
 
Equation (\ref{NUM21}) partially answers the question formulated prior to Eq.(\ref{NUM16}) and estimates the late-time value of the magnetic power spectrum. For the record let us now see what happens if Eq. (\ref{NUM14}) is not evaluated for $\tau= {\mathcal O}(\tau_{k})$
(i.e. when the relevant scale crosses the comoving Hubble radius) but as soon as radiation dominates i.e. for  $ \tau = {\mathcal O}(\tau_{1})$; obviously the power spectrum will not be larger but always much smaller than the one of Eq. (\ref{NUM21}) not because of the finite value of the conductivity (that only operates for $\tau > \tau_{k}$) but since $k \tau_{1} \ll 1$ when the relevant wavelength is larger than the Hubble radius. Let us then evaluate  Eqs. (\ref{NUM16})--(\ref{NUM18})  for $\tau\to \tau_{1}$;  the magnetic power spectrum, in this case, is given by:
\begin{equation}
{\mathcal Q}^{(em)}_{B}(k, \tau_{0}) =  \frac{9\,\, g_{1}^2\,\, \cos^2{\theta_{W}}}{128 \, \pi^2} \,\,\Omega_{R0} \,\, H_{0}^2 \, M_{P}^2\, {\mathcal A}_{{\mathcal R}}\, \,r_{T} \,\,\sin^2{k \tau_{1}},
\label{NUM22}
\end{equation}
where ${\mathcal Q}^{(em)}_{B}(k, \tau_{0})$ {\em now denotes the magnetic power spectrum obtained by evaluating the phases of oscillations not for $\tau = {\mathcal O}(\tau_{k})$ but rather for $\tau = {\mathcal O}(\tau_{1})$}. We actually use this different notation just for comparison since we regard as arbitrary the chain of arguments leading to ${\mathcal Q}^{(em)}_{B}(k,\tau_{0})$; for the record Eq. (\ref{NUM22}) corresponds to the strategy of Ref. \cite{bcf13}. The result of Eq. (\ref{NUM22}) is approximately, $|k \, \tau_{1}|^2 = {\mathcal O}(10^{-46})$ smaller than the one of Eq. (\ref{NUM21}) since, as already pointed out, $k \tau_{1} = {\mathcal O}(10^{-23})$; more specifically for the fiducial set of values of the concordance scenario we have \cite{TS1,TS2,TS3} 
\begin{equation}
k \tau_{1} = 5.75 \times 10^{-24} \biggl(\frac{k}{\mathrm{Mpc}^{-1}}\biggr) \biggl(\frac{r_{T}}{0.06}\biggr)^{-1/4}\, \biggl(\frac{{\mathcal A}_{{\mathcal R}}}{2.41 \times 10^{-9}}\biggr)^{-1/4} \biggl(\frac{h_{0}^2 \, \Omega_{R\,0}}{4.15 \times 10^{-5}}\biggr)^{-1/4}.
\label{NUM23}
\end{equation}
Inserting now Eq. (\ref{NUM23}) into Eq. (\ref{NUM22}) the explicit estimate 
of the magnetic power spectrum becomes
\begin{equation}
\frac{{\mathcal Q}^{(em)}_{B}(k, \tau_{0})}{\mathrm{nG}^2} = 1.12 \times 10^{-54} \biggl(\frac{r_{T}}{0.06}\biggr)^{1/2}\,\, \biggl(\frac{k}{\mathrm{Mpc}^{-1}}\biggr)^2 \,\,\biggl(\frac{g_{1}}{0.1}\biggr)^2\,\,
 \biggl(\frac{{\mathcal A}_{{\mathcal R}}}{2.41 \times 10^{-9}}\biggr)^{1/2}\,\, \biggl(\frac{h_{0}^2 \, \Omega_{R\,0}}{4.15 \times 10^{-5}}\biggr)^{1/2}.
\label{NUM24}
\end{equation}
Equation (\ref{NUM24}) is analog to the one of Ref. \cite{bcf13}; in that case the authors get $10^{-67} \mathrm{G}^2$ while we rather get (using the same argument based on Eq. (\ref{NUM22}))
${\mathcal Q}^{em}_{B}(k, \tau_{0}) = {\mathcal O}(10^{-72}) \,\, \mathrm{G}^2$. This quantitative disagreement comes from $\cos{\theta_{W}}^2$, from the value of $g_{1}$ and from a number of specific numerical factors (neglected in Ref. \cite{bcf13}) that further suppress the final magnetic field by roughly $5$ orders of magnitude.

Since the estimates of Eqs. (\ref{NUM20})--(\ref{NUM21}) depend on the post-inflationary expansion history we now consider how the scaling of the gauge fields is affected by the presence of an intermediate stage of expansion not necessarily associated with the radiation epoch. It has been actually suggested long ago that, under some very specific conditions, the presence of a post-inflationary stage may increase the magnetic power spectrum \cite{bcf8}. Even though a number of different possibilities can be considered \cite{SO3}, for the sake of conciseness we now illustrate the case of a single intermediate phase ranging between the end of inflation and the onset of radiation dominance. During the post-inflationary evolution  the expansion rate is controlled by a barotropic index $w$ and, for the fiducial values of  Eq. (\ref{NUM21}), the final result gets modified as:
 \begin{equation}
 {\mathcal P}^{(em)}_{B}(k,\tau_{0}) \,= 3.39\times 10^{- 8}\,\, \biggl(\frac{H_{r}}{H_{1}}\biggr)^{\alpha(w)}
\mathrm{nG}^2, \qquad \qquad \alpha(w) = \frac{2( 1 - 3 w)}{3(1+ w)},
 \label{NUM25}
 \end{equation}
where $H_{r}$ denotes the Hubble rate at the radiation dominance and, as usual, $H_{1}$ marks the end of the inflationary stage; by definition  $H_{r}/H_{1} < 1$ since the curvature scale at the beginning of inflation must always exceed the curvature scale at radiation dominance. The scale $H_{r}$ cannot be arbitrarily reduced since the intermediate stage(s) must not jeopardize the abundances of the light elements and, for this reason, we can generally require that $H_{r} > 10^{-44}\, M_{P}$ \cite{bcf8}. According to Eq. (\ref{NUM25}) there are three complementary physical possibilities characterized by the different ranges of $w$. When $w > 1/3$, the  power spectrum gets larger than in the case where $ w \to 1/3$ which is 
the one already discussed in Eq. (\ref{NUM21}): in this situation there is in fact a gain in the 
amplitude \cite{bcf8}. If $w < 1/3$ the magnetic power spectrum is instead 
further suppressed in comparison with the case $w \to 1/3$. Finally when $ w =1/3$ 
the result of Eq. (\ref{NUM21}) recovered and this is what must happen if the whole 
approach is consistent. The result of Eq. (\ref{NUM25})  tells that the magnetic power 
spectrum is {\em less diluted} when the dominant component of the background redshifts 
slower than radiation while it is {\em more suppressed} in the opposite case. 
As before the result of Eq. (\ref{NUM25}) can be compared with the expression derived from Eq. (\ref{NUM24}). We can then posit, by fiat, that 
in Eq. (\ref{NUM22}) the oscillating contribution should be evaluated in $\tau_{r}$ (and not in $\tau_{k}$).
For the typical values of the parameters previously introduced we then have  $k \tau_{r} = 10^{-23.24} (H_{1}/H_{r})$ and this result implies that ${\mathcal Q}^{(em)}_{B}(k, \tau_{0})$ is
\begin{equation}
{\mathcal Q}^{(em)}_{B}(k, \tau_{0}) = 1.12 \times 10^{-54} \biggl(\frac{H_{r}}{H_{1}}\biggr)^{\beta(w)}\,\,\,\mathrm{nG}^2,
\qquad\qquad \beta(w) = - \frac{1 + 9 w}{3 ( 1 + w)}.
\label{NUM26}
\end{equation}
Equation (\ref{NUM26}) stipulates that the final value of the magnetic power spectrum gets enhanced in spite of the value of the barotropic index $w$; this is rather bizarre especially because the enhancement even occurs when $w\to 1/3$ while we should expect that, in this situation, the result of a pure radiation phase 
is fully recovered. The ultimate reason of the result of Eq. (\ref{NUM26}) (which is incidentally the one 
suggested by Ref. \cite{bcf13}) is that the gauge power spectra have not been 
evaluated at the crossing time $\tau_{k}$ but much earlier without taking into account that, for 
$\tau \geq \tau_{k}$ the correct form of the physical power spectrum follows from Eq. (\ref{NUM18}).

\begin{figure}[!ht]
\centering
\includegraphics[height=8cm]{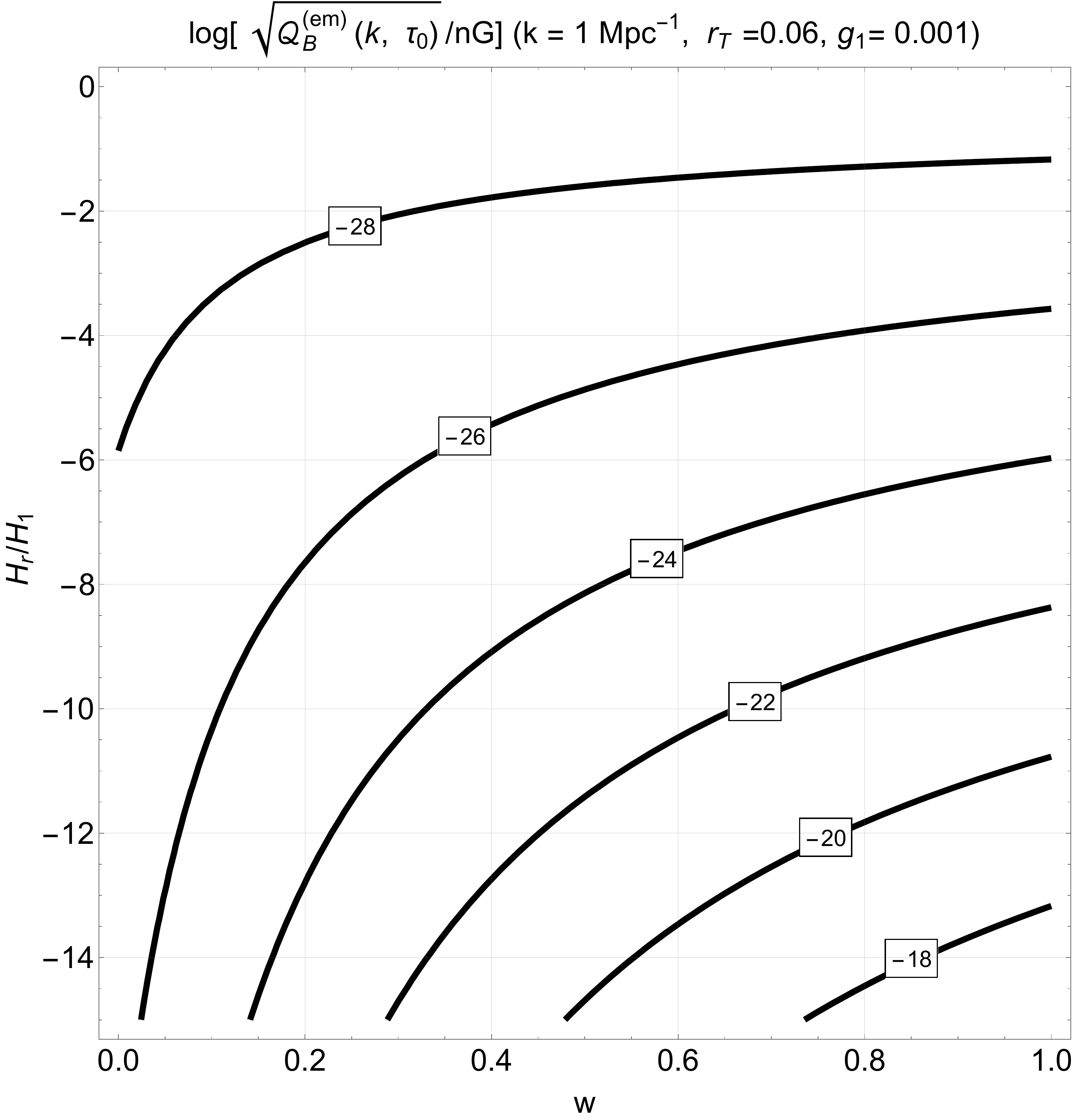}
\includegraphics[height=8cm]{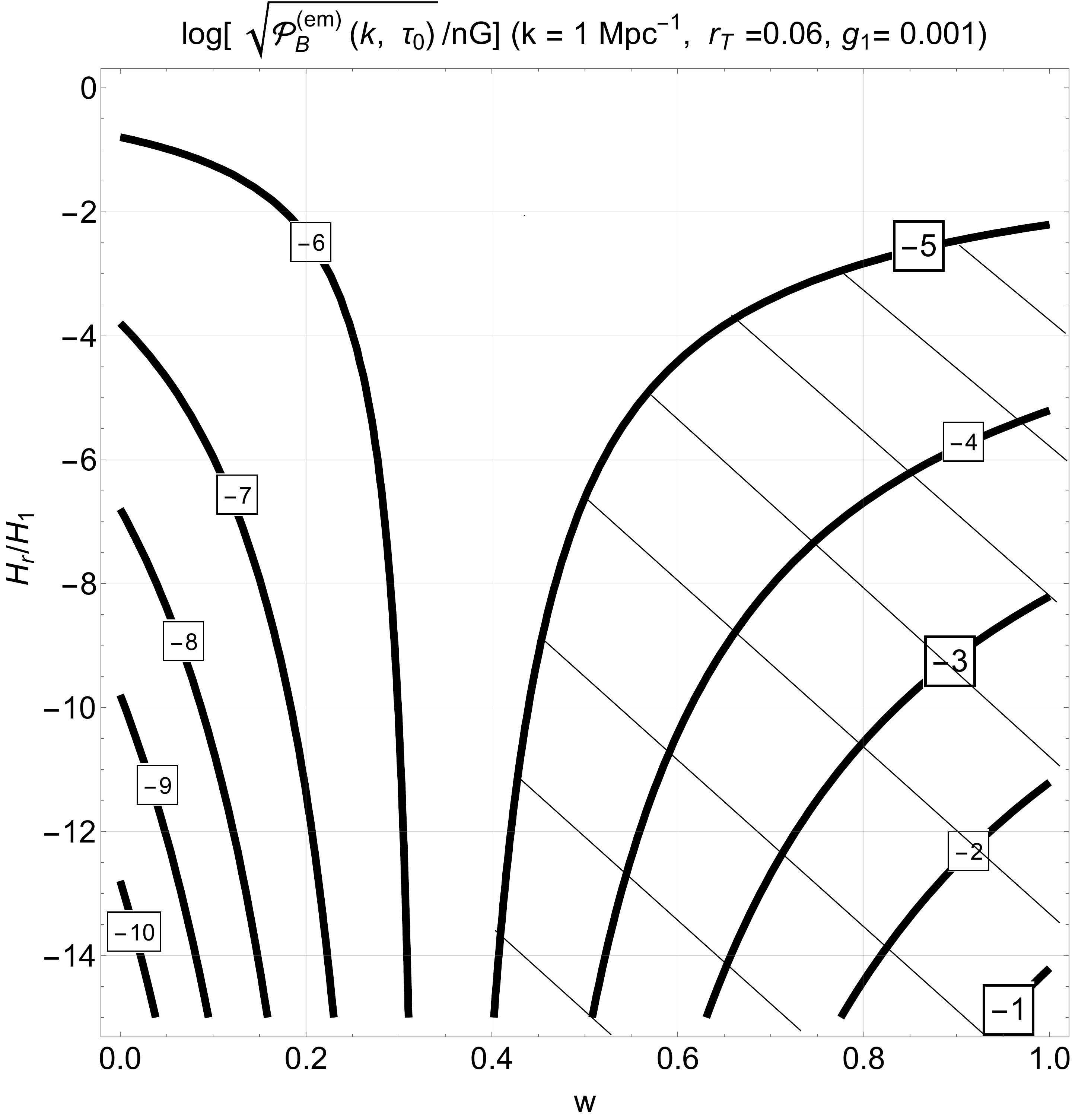}
\caption[a]{The common logarithms of the square roots of the magnetic 
power spectra are illustrated in the plane $(w, H_{r}/H_{1})$ for $r_{T}=0.06$ and for $k= 1\, \mathrm{Mpc}^{-1}$; note 
furthermore that $g_{1} = 0.001$. In the plot at the right we illustrate the result of Eq. (\ref{NUM25}) while at the left we considered the estimate Eq. (\ref{NUM26}). }
\label{FIG1}      
\end{figure}

For completeness in Fig. \ref{FIG1} we illustrated the results of Eqs. (\ref{NUM25})--(\ref{NUM26}) 
for a fiducial choice of the parameters. The common logarithm of the square roots of the magnetic power spectra (in nG units) is reported on the different contours of both plots.
In the shaded region (which is relevant for magnetogenesis considerations)  the power spectrum (\ref{NUM25}) is systematically larger than Eq. (\ref{NUM26}). Furthermore 
in the right plot the power spectrum is suppressed for $w<1/3$ while it is amplified 
when $w>1/3$. In spite of the previous comments (see Eq. (\ref{NUM26}) and discussion 
thereafter) it turns out that the results of the left plot of Fig. \ref{FIG1} are only 
marginally relevant for the magnetogenesis considerations. We actually know 
that the galaxy rotated roughly $30$ times since its origin and if we assume 
that the dynamo mechanism amplified the magnetic field of roughly one $e$-fold 
for each galactic rotation we would get that at the time of the gravitational collapse 
of the protogalaxy the initial magnetic field had to be, roughly, $10^{-19}$ G \cite{bcf18}. This 
number could be further reduced by taking into account the compressional 
amplification that would bring down the value by $4$ or $5$ orders of magnitude.
Even assuming the most optimistic coincidences we should have that, at least,
${\mathcal P}_{B}^{(em)} \geq 10^{-32}\, \mathrm{nG}$. A more realistic 
requirement would imply  ${\mathcal P}_{B}^{(em)} \geq 10^{-22}\, \mathrm{nG}$.
Both bounds are safely satisfied by the results of the right plot of Fig. \ref{FIG1}
while they are barely possible in the plot at the left. 

The results discussed so far have been obtained in the case when ${\mathcal F} = \pm {\mathcal H}$. Since this analysis can be generalized to the case ${\mathcal F} = \pm \gamma \, {\mathcal H}$ we now briefly examine, for illustration, the case of increasing gauge coupling (i.e. ${\mathcal F} = -\gamma {\mathcal H}$); in this case the spectral energy density during inflation is given by 
\begin{equation}
\Omega_{Y}(k, \tau) = \frac{1}{\rho_{crit}} \,\frac{d \langle \rho_{Y}\rangle}{d \ln{k}} = \frac{ 2 }{3 H^2 M_{P}^2 a^4} \biggl[ P_{E}(k,\tau) + P_{B}(k,\tau)\biggr],
\label{OMY}
\end{equation}
where $\rho_{crit}$ denotes the spectral energy density and $H$ is the inflationary Hubble rate. 
Equation (\ref{OMY}) can be written, in explicit terms, as:
\begin{equation}
\Omega_{Y}(k,\tau) = \frac{2}{3} \biggl(\frac{H}{M_{P}}\biggr)^2 \biggl[ {\mathcal C}(|\gamma -1/2|) |k\tau|^{5 - |2 \gamma-1|} + {\mathcal C}(\gamma +1/2) |k\tau|^{4 - 2 \gamma} \biggr],
\label{OMY2}
\end{equation}
where ${\mathcal C}(x) = 2^{2x -3} \, \Gamma^2(x)/\pi^3$. From Eq. (\ref{OMY2}) it follows 
that $\gamma \leq 2$ to avoid that the spectral energy density eventually becomes 
overcritical. After the end of inflation the spectral energy density can be expressed as
\begin{equation}
\Omega_{Y}(k, \tau) = \frac{2 \, H_{1}^4}{3 \, H^2 \, M_{P}^2} \,\, \biggl(\frac{a_{1}}{a}\biggr)^4 {\mathcal C}(\gamma +1/2) \,\, \biggl(\frac{k}{a_{1} \, H_{1}}\biggr)^{4 - 2 \gamma}.
\label{OMY3}
\end{equation}
If we now evaluate Eq. (\ref{OMY3}) for $\tau\to \tau_{r}$ and require that $\Omega_{Y}(k, \tau_{r}) < 10^{-6}$ 
we get a bound in the plane ($H_{r}/H_{1}$, $w$) which is incidentally satisfied in the dashed area 
of Fig \ref{FIG1}.

We can therefore summarize the main findings of this analysis as follows.
The gauge fields do not experience any unexpected growth in a conformally flat cosmological background. When the Fourier mode gets of the order of the (comoving) Hubble radius the peculiar phases of the gauge power spectra are of order $1$ and do not entail an increase of the final magnetic field. 
The present results show that the Sakharov phases must be evaluated when 
the wavelengths relevant for magnetogenesis cross the Hubble radius 
and then corrected for the effect of the conductivity when the corresponding scales are inside the horizon. For the typical parameters of the concordance scenario the crossing happens slightly before matter-radiation equality it would be arbitrary to evaluate the gauge power spectra well before horizon crossing or simply at the end of inflation. In spite of recent claims the post-inflationary stages different from the radiation phase {\em do not always improve the magnetogenesis constraints}. Conversely, as suggested long ago, only the presence of a sufficiently long stiff epoch  improves the situation since, in this case, the expansion rate is slower than radiation. If the expansion rate 
is instead faster than radiation the final values of the magnetic power spectra are comparatively smaller 
than in the case of a radiation-dominated evolution. The previous conclusions are also consistent 
with the scaling of the gauge fields dictated by  the covariant conservation of the corresponding  energy momentum tensor since the hypermagnetic power spectra remain approximately constant in comparison with the background energy density when radiation is the dominant component of the post-inflationary plasma. 

I wish to thank T. Basaglia,  A. Gentil-Beccot, S. Rohr and the whole CERN Scientific Information 
Service for their help during the preparation of this manuscript.

\end{document}